 \documentclass[conference, a4paper, 10pt]{IEEEtran}
\usepackage{geometry}
    
\usepackage{color}
\usepackage{xcolor}
\usepackage{cuted}
\usepackage{etoolbox}
\usepackage{mathrsfs}
\usepackage{amssymb}
\usepackage[mathcal]{euscript}
\usepackage{cite}
\usepackage{graphicx}
\usepackage{psfrag}
\usepackage{subfigure}
\usepackage{url}
\usepackage{stfloats}
\usepackage{amsmath}
\usepackage{float}
\usepackage{array}

\usepackage{amsthm}
\usepackage{booktabs} %±ížñ
\usepackage{algorithmic} %format of the algorithm
\usepackage[linesnumbered,ruled,vlined]{algorithm2e}
\usepackage{color}
\usepackage{cases}
\usepackage{bm}
\usepackage{amsmath,amsfonts,amssymb,amsbsy,paralist,ifthen}

\usepackage{colortbl}
\usepackage{arydshln}
\usepackage{multirow}
\usepackage{multicol}

\usepackage{color}

\ifCLASSINFOpdf
\else
\fi
\begin{document}

% \title{Sum Rate Maximization for Semi-Self Sensing HRIS-Aided THz Integrated Sensing and Communications\\
% \title{Sum Rate Enhancement using Machine Learning for Simultaneous ISAC\textendash capable Transmission and Sensing\textendash \textcolor{blue}{Augmented} HRIS\textendash Enabled THz Communications\\
\title{Sum Rate Enhancement using Machine Learning for Semi-Self Sensing Hybrid RIS-Enabled ISAC in THz Bands\\
{\footnotesize}
% \thanks{Identify applicable funding agency here. If none, delete this.}
}

\author{Sara Farrag Mobarak\IEEEauthorrefmark{2}, Tingnan Bao\IEEEauthorrefmark{2},  Melike Erol-Kantarci\IEEEauthorrefmark{2}\\
\IEEEauthorblockA{\IEEEauthorrefmark{2}\textit{School of Electrical Engineering and Computer Science, University of Ottawa, Ottawa, Canada}}
\IEEEauthorblockA{Emails:\{smobarak, tbao, melike.erolkantarci\}@uottawa.ca}}
\maketitle
\begin{abstract}
This paper proposes a novel semi- self sensing hybrid reconfigurable intelligent surface (SS- HRIS) in terahertz (THz) bands, where the RIS is equipped with reflecting elements divided between passive and active elements in addition to sensing elements. SS-HRIS along with integrated sensing and communications (ISAC) can help to mitigate the multipath attenuation that is abundant in THz bands. In our proposed scheme, sensors are configured at the SS-HRIS to receive the radar echo signal from a target. A joint base station (BS) 
 beamforming and HRIS precoding matrix optimization problem is proposed to maximize the sum rate of communication users while maintaining satisfactory sensing performance measured by the Cramér- Rao bound (CRB) for estimating the direction of angles of arrival (AoA) of the echo signal and thermal noise at the target. The CRB expression is first derived and the sum rate maximization problem is formulated subject to communication and sensing performance constraints. To solve the complex non- convex optimization problem, deep deterministic policy gradient (DDPG)- based deep reinforcement learning (DRL) algorithm is proposed, where the reward function, the action space and the state space are modeled. Simulation results show that the proposed DDPG- based DRL algorithm converges well and achieves better performance than several baselines, such as the soft actor–critic (SAC), proximal policy optimization (PPO), greedy algorithm and random BS beamforming and HRIS precoding matrix schemes. Moreover, it demonstrates that adopting HRIS significantly enhances the achievable sum rate compared to passive RIS and random BS beamforming and HRIS precoding matrix schemes. 

\end{abstract}

%\begin{IEEEkeywords}
%Wireless communication, Integrated Sensing and Communication (ISAC), Hybrid RIS, Sensing RIS, Sum Rate, Cramer Rao Bound (CRB), TD3, Beamforming, THz. \end{IEEEkeywords}

%%%%%% 5 keywords %%%%%%%
\begin{IEEEkeywords}
Integrated Sensing and Communication (ISAC), Hybrid Reconfigurable Intelligent Surface (HRIS), \textcolor{black}{sensing RIS}, DDPG, THz. \end{IEEEkeywords}

\section{Introduction}
\textcolor{black}{Incorporating sensing capabilities into wireless communication networks has recently emerged as an important feature in the advancement of beyond fifth-generation (B5G) as well as the sixth- generation (6G) networks \cite{lu2024integrated}. To reduce the hardware costs, decrease power consumption and boost the spectral efficiency, sharing the same time-frequency resources and hardware platform between radar and communication has recently attracted research interest and 
attention from both the industry \cite{zhao2024performance}. Furthermore, integrated sensing and communication (ISAC) allows wireless networks to gather sensory data from the surroundings, thereby contributing to the development of smart environment\textendash aware technologies \cite{cui2023integrated}}. 
\textcolor{black}{In this evolving ISAC landscape, conventional RISs have been employed to assist communication by enhancing data transmission. However, sensing\textendash augmented RIS (SA-RIS) introduces a new role where the RIS can be deployed to assist ISAC by facilitating line-of-sight (LOS) paths for sensing tasks handled by the BS alone 
\cite{saikia2024hybrid,gan2024bayesian}.
The SA\textendash RIS simply reflects the probing signals generated by the BS, allowing the BS to carry out target detection or environmental sensing with enhanced reach and accuracy. Moving beyond this reflective assistance, a more advanced concept, known as semi-self sensing RIS (SS\textendash RIS), has recently been introduced to reduce the reliance on the BS for sensing tasks. Unlike SA\textendash RIS, SS\textendash RIS incorporates a mix of reflecting and sensing elements \cite{shao2022target}. While the SS-RIS still depends on the BS to generate probing signals, it can directly receive echo signals from targets, allowing for basic radar sensing \cite{lyu2024crb}. This semi-autonomous design reduces signal degradation associated with multi-hop paths (e.g., BS $\rightarrow$ RIS $\rightarrow$ target $\rightarrow$ RIS $\rightarrow$ BS), as it only requires a two-hop reflection (i.e., BS $\rightarrow$ RIS reflecting elements $\rightarrow$ target $\rightarrow$ RIS sensing elements)  \cite{shao2024intelligent}.} 

The field of \textcolor{black}{SS\textendash} RIS-assisted ISAC is still in its nascent stages 
% \notegreen{the title says enabled now it became assisted ?? you should be consistent with wording. please check all the paper. assisted is a better word for your conext} \notered{I feel like you are using multiple names for the same thing. let's decide on one and use it. so, my undersatnding is RIS has radar sensors so RIS becomes a ISAC platform. This can be ISAC-enabled RIS maybe?, do we need semi-self or hybrid in that case? are these well known names to RIS community?}
% \textcolor{blue}{I guess I have to give the RIS a name to indicate that the RIS itself is having sensors, not only implemented to provide LOS in ISAC "ISAC\textendash enabled RIS" where the BS is the one responsible for sensing}
, with only a handful of studies available in the literature \cite{shao2022target,lyu2024crb,he2023joint,shao2024target }. Inspired by \cite{shao2022target}, which is the first work to propose the concept of \textcolor{black}{SS\textendash } RIS, the authors in \cite{lyu2024crb} explored a \textcolor{black}{millimeter\textendash wave} (mmWave) ISAC system, where an \textcolor{black}{SS\textendash} passive RIS is deployed to provide connectivity between the BS, communication users, and targets. In their work, a joint optimization problem is formulated on hybrid BS beamforming and RIS phase shifts to minimize the CRB, while guaranteeing good communication performance, evaluated by the achievable data rate. In \cite{he2023joint}, the authors investigated joint channel and AoA estimation
in an \textcolor{black}{SS\textendash} RIS- unmanned aerial vehicle (UAV) network, where the effect of the \textcolor{black}{SS\textendash} RIS power splitting coefficient on the estimations of the
individual channels and the AoAs of the LOS path of the UAV-RIS
link is analyzed. \\
% \textcolor{purple}{Lastly, the work in \cite{shao2024target} proposed a secure wireless sensing approach aimed at simultaneously enhancing target detection for legitimate radar station (LRS)
% while securing it from an unauthorized radar station (URS) by deploying the target-mounted SS\textendash RIS. In their work, optimization problems are formulated to design the reflecting phase shifts at RIS to maximize the received signal power at the LRS while keeping that, at the URS, below a certain level, for both cases of short and long\textendash term RIS operations with different dynamic reflection capabilities.}\\
% a protocol for the target- mounted \textcolor{blue}{self}\textendash sensing RIS and designed the RIS phase shifts is designed to maximize the reflected signal power toward the LRS.
% \notered{you might need to revise the above text based on what naming we use?}\\
Additionally, since passive RIS is only capable of reflecting the incident signal with no amplification gain introduced and the capacity gain provided by passive RIS is limited due to multiplicative fading effect, HRIS is proposed. Here, some of the RIS reflecting elements are active (e.g. amplification gain introduced) and the rest are passive. This combination of active and passive RIS elements is proven to be the optimal selection to address the trade\textendash off the system performance and hardware costs. Hence, it is a promising approach to deploy HRIS in the ISAC systems \cite{hao2024joint}.\\
Despite several research studies, analyzing the sum rate of the ISAC downlink system through an \textcolor{black}{SS\textendash} sensing HRIS has not been explored, in particular considering the THz band and the sensing performance is guaranteed.
%Tingnan's comment 3:
% \textcolor{orange}{If you will use this term "semi–self sensing RIS", please update "semi–self sensing
% HRIS" into "semi–self sensing RIS". Otherwise, the author will be confused for a new definition of RIS. Just keep it use "semi–self sensing RIS" in the whole paper. Besides, you can also consider to abbreviate this term "semi–self sensing RIS" into "SS-RIS" for short. Otherwise, it will read a little bit redundancy.} 
In this paper, we specifically tackle this problem by considering an \textcolor{black}{SS\textendash} HRIS, equipped with both reflective (passive and active) elements and sensing elements, in an ISAC downlink network to establish LOS between the BS and the communication users (and a target) in the THz band. A DDPG\textendash based DRL algorithm is employed to jointly optimize the BS beamforming and the HRIS precoding matrix.
Our main contributions can be summarized as follows:

\begin{itemize}
  \item A joint BS beamforming and \textcolor{black}{SS\textendash} HRIS precoding matrix optimization scheme is proposed to maximize the sum rate of the scenario under consideration given the constraints of the HRIS, sensing performance measured by CRB \textcolor{black}{and thermal noise sensed at the target} and limited power budgets of both the BS and HRIS. 
  \item Given the formulated sum rate maximization problem is non-convex non-trivial one, due to the non-convexity of both the objective function and the constraints, we reformulate the problem within the framework of DRL. The 
  DDPG algorithm is utilized to derive the feasible solutions for $\textbf{W}$ and $\mathbf{\Phi}$ as the outputs of the DRL neural network.
  \item Simulation results demonstrated that DDPG readily outperforms other benchmark algorithms, such as PPO, SAC, Greedy and random algorithms in terms of the sum rate. Moreover, HRIS is found to surpass the passive RIS and random schemes thanks to its provided substantial gains.
\end{itemize}
% \textcolor{red}{Notations: $[.]^{-1}$, $[.]^{T}$ and $[.]^{H}$ denote the inverse, transpose and Hermitian operations, respectively. ||.|| denotes the Frobenius norm and $\otimes$ denotes the Kronecker product. $\mathbb{C}$, $\mathbb{R}$ and $\mathbb{R}_{+}$ denote the set of complex, real and positive real numbers, respectively. $\boldsymbol{I_{N}}$ and $\boldsymbol{0_{N}}$ denote an identity $N\times N$ matrix and $1\times N$ zero vector, respectively.}

\section{System and Channel Model}
\subsection{System Model}
Consider a downlink ISAC model operating in THz band, where a BS transmits signals to serve $K$ single\textendash antenna communication users in addition to probing signals to \textcolor{black}{sense} a single target. 
% \notered{what does serve a single target mean? Please explain the scenario better here. what is sensing doing what are users aiming, etc?}
% \textcolor{blue}{I am sorry, I meant "sense" not "serve". I also used the same system model here \cite{lyu2024crb}, Fig. 1}, as illustrated in Fig \ref{SystemModel}. 
% Tingnan comment 1:
% \textcolor{orange}{I think we can use "a single target sensor" instead of "a single target" for better understanding and also update it over the whole paper.} 
%My reply: but it is a single target, not single target sensor
Without loss of generality, the direct links between the BS and the users, as well as that between the BS and the target, are assumed to be unattainable due to high loss caused by obstacles and long transmission distance. Consequently, an \textcolor{black}{SS\textendash} HRIS is deployed, equipped with both reflecting \textcolor{black}{(passive and active) elements} as well as sensing elements, to establish LOS between the BS and the users (and the target). The BS is assumed to be equipped with a uniform plannar array (UPA) consisting of $M = M_{x} M_{z}$ transmit antenna elements deployed in the $x-z$ plane, the \textcolor{black}{SS\textendash} HRIS can be also treated as a UPA consisting of $N = N_{y} N_{z}$ reflecting elements and $N_s = N_{s_{y}}N_{s_{z}}$ sensing elements deployed in the $y-z$ plane.

%The channels between the communication nodes in the network are considered to be THz channel\textcolor{black}{s}. It is assumed that the  channels between the nodes are known or at least can be perfectly estimated. The description of this channel is provided in the next subsection.
\begin{figure}[ht]
\centering
\vspace{-.1em}
\includegraphics[height=5cm,width=5cm]{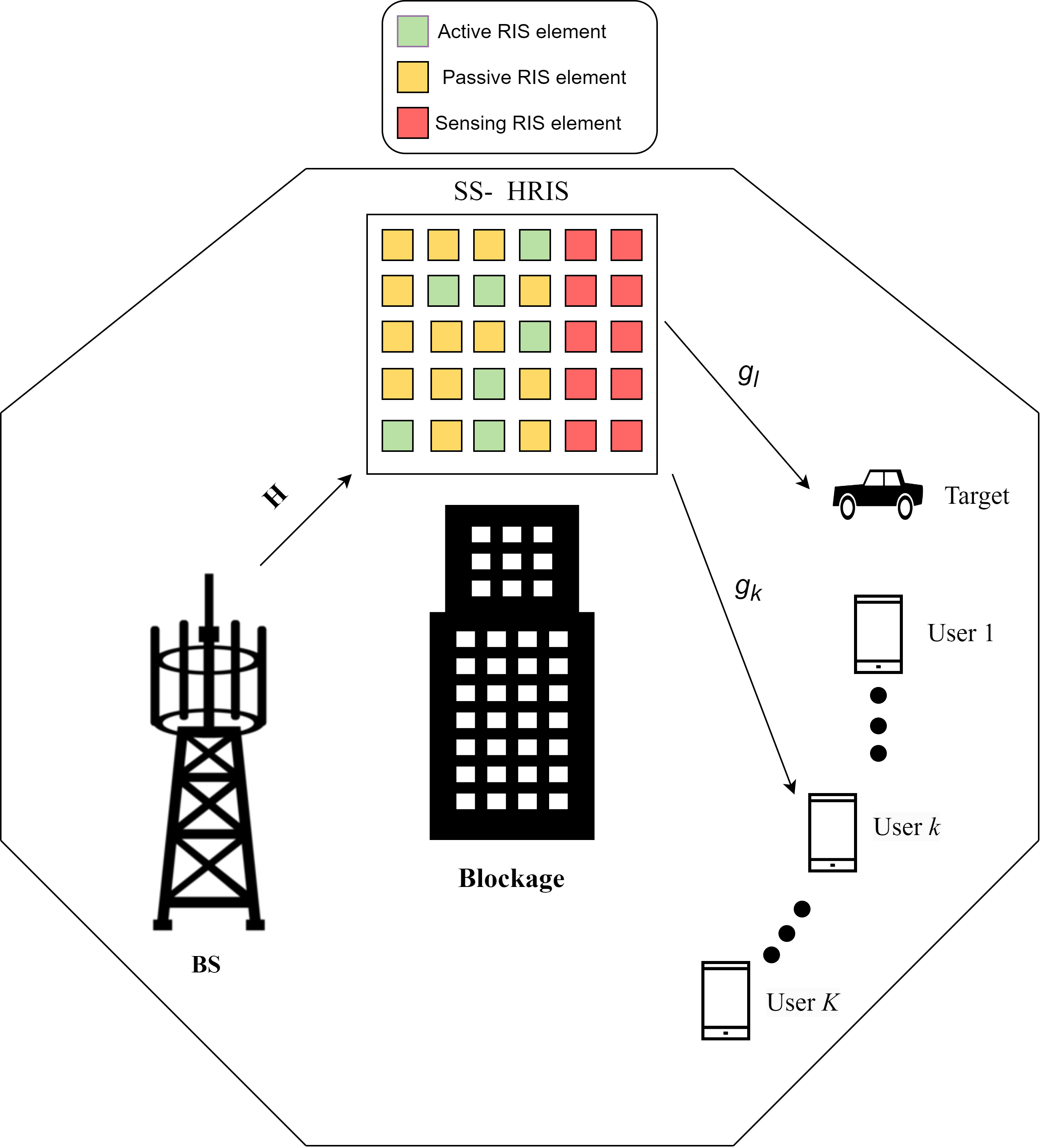}
\caption{Scenario considering an \textcolor{black}{SS\textendash} HRIS that is deployed to assist the communication link between the BS and $K$ communication users and a target.}\label{SystemModel}
\end{figure}

\subsection{Channel Model}
% In this sub-section, the THz channel model is illustrated. It was derived using THz wave atmospheric transmission attenuation model in addition to  water vapour absorption which primarily  characterize the THz band.
Consider the THz wave transmission attenuation model and water vapour absorption that primarily characterize the THz band \cite{zhang2020energy,boulogeorgos2018performance}, the THz channel matrix between the BS and the RIS reflecting elements is expressed as
\begin{equation}
\label{H_eqn}
\textbf{H} = \sqrt{\frac{N M}{PL(f,d)}} \bm{\alpha}_{r}(\psi^{r},\omega^{r}) \bm{\alpha}_{t}^{H}(\psi^{t},\omega^{t}),
\end{equation}
where $PL(f,d)$  represents the pathloss experienced at frequency $f$ after propagating a distance $d$ and is given as
\vspace{-2mm}
\begin{equation}
\label{channel_model}
PL{(f,d)} = 
 \Big(\frac{4\pi f d}{C}\Big)^{2}  e^{k(f)d},
\end{equation}
where $C$ and $k(f)$ represent the speed of light in free space and the frequency dependent medium absorption factor, respectively. Also, $ \bm{\alpha}_{\textcolor{black}{r}}(\psi^{\textcolor{black}{r}}, \omega^{\textcolor{black}{r}})$ and $ \bm{\alpha}_{\textcolor{black}{t}}(\psi^{t}, \omega^{t})$ in \textcolor{black}{(\ref{H_eqn})} represent RIS and BS steering vectors, which are expressed respectively as 
\begin{subequations}
\begin{align}
    \bm{\alpha}_{r}(\psi^{r}, \omega^{r}) &= \frac{1}{\sqrt{N}}
    \begin{bmatrix}
        1, \ldots, e^{-j \frac{2\pi}{\lambda} (N_{y}-1) \sin \psi^{r} \sin \omega^{r} d_r}
    \end{bmatrix}^{T} \notag \\ 
    &\otimes
    \begin{bmatrix}
        1, \ldots, e^{-j \frac{2\pi}{\lambda} (N_{z}-1) \cos \omega^{r} d_r}
    \end{bmatrix}^{T}, \tag{3a} \\[2mm]
\bm{\alpha}_{t}(\psi^{t} , \omega^{t}) &= \frac{1}{\sqrt{M}}
\begin{bmatrix}
1, \ldots, e^{-j \frac{2\pi}{\lambda} (M_x - 1) \cos \psi^{t} \sin \omega^{t} d_b}
\end{bmatrix}^T \notag \\ 
&\otimes 
\begin{bmatrix}
1, \ldots, e^{-j \frac{2\pi}{\lambda} (M_z - 1) \cos \omega d_b}
\end{bmatrix}^T. \tag{3b}
\end{align}
\end{subequations}
% \vspace{-1mm}
%  \begin{align}
%  \begin{aligned}
%     \mathbf{\alpha}(\psi, \omega) & = \frac{1}{\sqrt{N_s}}
%     \begin{bmatrix}
%         1, \ldots, e^{-j \frac{2\pi}{\lambda} (N_{sy}-1) \sin \psi \sin \omega d_r}
%     \end{bmatrix}^{T}
%     \\& \otimes
%     \begin{bmatrix}
%         1, \ldots, e^{-j \frac{2\pi}{\lambda} (N_{sz}-1) \cos \phi d_r}
%     \end{bmatrix}^{T}
%     \end{aligned}
% \end{align}
%  \begin{align}
%  \begin{aligned}
% \mathbf{\beta}(\psi, \omega) & = \frac{1}{\sqrt{N}}
% \begin{bmatrix}
% 1, \ldots, e^{-j \frac{2\pi}{\lambda} (N_y - 1) \sin \psi \sin \omega d_b}
% \end{bmatrix}^T 
% \\& \otimes 
% \begin{bmatrix}
% 1, \ldots, e^{-j \frac{2\pi}{\lambda} (N_z - 1) \cos \omega d_b}
% \end{bmatrix}^T
% \end{aligned}
% \end{align}
where $\psi^{r}$, $\omega^{r}$, $\psi^{t}$, $\omega^{t}$ and $d_r$ and $d_b$ denote azimuth/elevation angles of arrival/departure (AoA/AoD), the element spacing of RIS elements and transmit antenna elements at the BS. The channel from the RIS to user $k$ can be modeled in a similar manner, with only transmit array response due to a single receive antenna at users.
% \vspace{-2mm}
% \notered{in the above text we lost all the connection to ISAC. explain which parts are related to isac}
% \textcolor{violet}{This part is not intended to be related to sensing. I was only explaining the channel model (BS $\rightarrow$ RIS), (RIS $\rightarrow$ user (or target)), like normal channels in communications system not having ISAC} 
%Tingnan's comment 4: 
% \textcolor{orange}{We can reorganize the architecture of this session to avoid the mis-understanding, including B. Communication Model, C. Sensing Model D. Problem Formulation.}
% $PL(f,d)$ accounts for spreading loss $L_{sl}(f,d)$ as well as molecular absorption $L_{mal}(f,d)$ that distinguishes the THz frequency band. \textcolor{magenta}{The spreading loss $L_{sl}(f,d)$ is resulting from the expansion of the electromagnetic wave when it propagates through unlike mediums.} On the other hand, \textcolor{magenta}{the molecular absorption $L_{mal}(f,d)$ is a consequent of the collisions between water molecules  and/or atmospheric gas. }Detailed study for the effect of atmospheric attenuation was carried out in \cite{boulogeorgos2018distance}.
% $PL(f,d)$ accounts for spreading loss as well as molecular absorption that distinguishes the THz frequency band. 
% Detailed study for the effect of atmospheric attenuation was carried out in \cite{boulogeorgos2018distance}.
% The channel coefficient $\textbf{H}_{THz}$ follows zero-mean complex Gaussian distribution, having a variance that models free space path and molecular absorption gain. From \cite{boulogeorgos2018distance} (Eqn. (2), (3) and (5)),
\section{Mathematical Formulation}
\subsection{\textcolor{black}{Metrics for Communications}}
The received signal at each communication user $k$ can be expressed as
\begin{align}
\begin{aligned}
\label{ri}
y_{k}(t) = & \underbrace{ \boldsymbol{g}_{k}^{H} \boldsymbol{\Phi} \boldsymbol{H} \boldsymbol{w}_{k} s_{k} (t)}_{\text {Desired Signal}} +  \sum_{\substack{j = 1\\
                  j \ne i
                  }}^{K} \underbrace{ \boldsymbol{g}_{k}^{H} \boldsymbol{\Phi} \boldsymbol{H} \boldsymbol{w}_{j} s_{j}(t)}_{\text{ Inter- user Interference}}\\& + \underbrace{\boldsymbol{g}_{k}^{H} \boldsymbol{A}\boldsymbol{\Phi} \boldsymbol{n}_{a}}_{\text {Dynamic Noise}} + \underbrace{\boldsymbol{g}_{l}^{H} \boldsymbol{\Phi} \boldsymbol{H} \boldsymbol{w}_{K+1} s_{K+1}(t)}_{\text {Sensing Interference}} + \underbrace{n_{o}}_{ \text{AWGN}},
\end{aligned}
\end{align}
where $\boldsymbol{g}_{k} \in \mathbb{C}^{N\times 1}$ denotes the channel vector between the HRIS and communication user $k$,
$\boldsymbol{\Phi} \in \mathbb{C}^{N\times N} $ denotes the precoding matrix of the  HRIS coefficients (i.e., phase shifts and amplitudes), where $\boldsymbol{\Phi}$ = diag  [$\eta_{1} e^{j\theta_{1}},...,\eta_{N} e^{j\theta_{N}}]^{H}$ and $\theta_{n} \in [\ 0, 2 \pi$) and $\boldsymbol{H}\in\mathbb{C}^{N\times M}$ is the channel matrix between the UPA of the BS and the HRIS. We define $\boldsymbol{W} = [\boldsymbol{w}_{1}, ...., \boldsymbol{w}_{K+1}]$, where $\boldsymbol{w}_{k} \in \mathbb{C}^{M\times 1}$ is the BS beamforming for user $k$, $\boldsymbol{s}(t) \in \mathbb{C}^{K+1\times 1}$ represents the transmit data symbol that consists of $K$ data streams for serving the $K$ communication users and one stream for the sensing target such as $\boldsymbol{s}(t) = [\underbrace{s_{1}(t), ..., s_{K}(t)}_{Communication}, \underbrace{s_{K+1} (t)}_{Sensing}]^{T}$. Also, $\boldsymbol{A}$ is defined as a selection matrix represented as a diagonal matrix with a total of $q$ ones in its diagonal, where $q$ represents the number of active elements in the HRIS which are randomly assigned, $\boldsymbol{n}_{a} \sim CN (0, \sigma_{a}^{2} \boldsymbol{I_{N}})$ represents the dynamic noise generated by the active elements, which is related to the input noise as well as inherent device noise of active RIS elements \cite{zhang2021active}, and $n_{o} $ denotes the additive white Gaussian noise (AWGN) which has zero mean and variance $\sigma_{o}^{2}$.

According to (\ref{ri}), the achieved SINR at the communication user $k$ can be given by 
 % \small
\begin{align}
\begin{aligned}
\label{SINR}
 &\gamma_{k} =\\ &\frac{|  \boldsymbol{g}_{k}^{H} \boldsymbol{\Phi} \boldsymbol{H} \boldsymbol{w}_{k}|^{2}}{ \sum_{\substack{j = 1\\
                  j \ne i
                  }}^{K} | \boldsymbol{g}_{k}^{H} \boldsymbol{\Phi} \boldsymbol{H} \boldsymbol{w}_{j}|^{2} + |\boldsymbol{g}_{l}^{H} \boldsymbol{\Phi}  \boldsymbol{H} \boldsymbol{w}_{K+1}|^{2} + ||\boldsymbol{g}_{k}^{H} \boldsymbol{A}\boldsymbol{\Phi}||^{2} \sigma_{a}^{2}+ \sigma_{o}^{2}}.
      \end{aligned}
\end{align}
% \footnotesize
Accordingly, the achievable data rate of user $k$ is expressed as
\begin{equation}
    R_{k} = \log_{2} (1 + \gamma_{k}).
\end{equation}
\subsection{\textcolor{black}{Metrics for Sensing}}
As for the target, we assume it is in the far field of the BS and the \textcolor{black}{SS\textendash} HRIS so that it can be viewed as a point\textendash like target. As such, the steering vectors of the HRIS sensing elements and reflecting elements can be expressed respectively as
\begin{subequations}
\begin{align}
\bm{\alpha}(\psi, \omega) &= \frac{1}{\sqrt{N_s}} 
\left[ 
\begin{array}{c}
1, \dots, e^{-j\frac{2\pi}{\lambda}(N_{sy}-1)\sin\psi\sin\omega d_y}
\end{array}
\right]^T  \notag \\
&\otimes
\left[ 
\begin{array}{c}
1, \dots, e^{-j\frac{2\pi}{\lambda}(N_{sz}-1)\cos\omega d_z}
\end{array}
\right]^T, \\
\bm{\beta}(\psi, \omega) &= \frac{1}{\sqrt{N}} 
\left[ 
\begin{array}{c}
1, \dots, e^{-j\frac{2\pi}{\lambda}(N_{y}-1)\sin\psi\sin\omega d_y}
\end{array}
\right]^T \notag \\
&\otimes
\left[ 
\begin{array}{c}
1, \dots, e^{-j\frac{2\pi}{\lambda}(N_{z}-1)\cos\omega d_z}
\end{array}
\right]^T,
\end{align}
\end{subequations}
where $\lambda$ denotes the carrier wavelength, and $d_y$ and $d_z$ denote the horizontal and vertical adjacent HRIS element spaces, respectively, and $\psi$ and $\omega$ are the azimuth and elevation angles.\\
In addition, the received echo signal from the target at the sensors of the HRIS can be expressed as
\begin{align}
\label{Ytarget}
\begin{aligned}
        \boldsymbol{Y}_{target} = &\rho \boldsymbol{\alpha}(\psi, \omega) \boldsymbol{\beta}^{T}(\psi, \omega) \boldsymbol{\Phi}\boldsymbol{H}\boldsymbol{X} \\+& \rho \boldsymbol{\alpha}(\psi, \omega) \boldsymbol{\beta}^{T}(\psi, \omega) \boldsymbol{\Phi}\boldsymbol{N}_{a} + \boldsymbol{N_{o}},
        \end{aligned}
\end{align}
% where $\rho$ represents a complex constant counting for the round\textendash trip pathloss as well as the radar cross section (RCS) at the target, $d_{y}$ and $d_{z}$ represent the horizontal and vertical adjacent RIS elements spacing, $\textbf{N}_{a}$ denotes the HRIS dynamic noise with each entry being $\sigma_{a}^2$ and $\textbf{N}_{o}$ denotes the AWGN with each entry being $\sigma_{o}^2$.  Assume that the fluctuations of the RCS are  slow and the round\textendash trip sensing channel is unchanged during the transmission of $T$ the communication and sensing symbols, where the Swerling\textendash I model is applicable \cite{eaves2012principles}. From eqn. (\ref{Ytarget}), it is noted that the transmitted signal, received signal and the AWGN are stacked as $\textit{\textbf{X}}= [\textit{\textbf{x}}(1),…,\textit{\textbf{x}}(T)]$, $\textit{\textbf{Y}}_{target} =[\textit{\textbf{y}}_{target (1)},…,\textit{\textbf{y}}_{target(T)}]$ and $\textit{\textbf{N}}_o =[\textit{\textbf{n}}_o (1),…,\textit{\textbf{n}}_o(T)]$, respectively, where $T$ also represents the radar dwell time. When $T$ is very large, the covariance matrix of the transmit signal $\textbf{x}(t)$ can be written as
where $\rho$ represents a complex constant counting for the round\textendash trip pathloss as well as the radar cross section (RCS) at the target, $d_{y}$ and $d_{z}$ represent the horizontal and vertical adjacent RIS elements spacing, $\mathbf{N}_{a}$ denotes the HRIS dynamic noise with each entry being $\sigma_{a}^2$, and $\mathbf{N}_{o}$ denotes the AWGN with each entry being $\sigma_{o}^2$. Assume that the fluctuations of the RCS are slow and the round\textendash trip sensing channel is unchanged during the transmission of $T$ communication and sensing symbols, where the Swerling\textendash I model is applicable \cite{eaves2012principles}. From  (\ref{Ytarget}), it is noted that the transmitted signal, received signal, and the AWGN are stacked as $\mathit{\mathbf{X}}= [\mathit{\mathbf{x}}(1), \ldots, \mathit{\mathbf{x}}(T)]$, $\mathit{\mathbf{Y}}_{target} =[\mathit{\mathbf{y}}_{target}(1), \ldots, \mathit{\mathbf{y}}_{target}(T)]$ and $\mathit{\mathbf{N}}_o =[\mathit{\mathbf{n}}_o(1), \ldots, \mathit{\mathbf{n}}_o(T)]$, respectively, where $T$ also represents the radar dwell time. When $T$ is very large, the covariance matrix of the transmit signal $\mathbf{x}(t)$ can be written as:
\begin{equation}
    \textbf{R}_{\mathbf{x}} = \mathbb{E}\{\mathbf{x}(t)\mathbf{x}^{H}(t)\} = \textbf{W}\textbf{W}^{H} \approx \frac{1}{T} \textbf{X} \textbf{X}^{H}.
\end{equation}
As such, $\textbf{Y}_{target}$ is vectorized so that the following holds \cite{lyu2024crb}
\begin{equation}
\textbf{y}_{target} = vec(\mathbf{Y}_{target} )  = \textbf{b} + \textbf{n}_{o},
\end{equation}
where
%\textcolor{red}{\rho \boldsymbol{\alpha}(\psi, \omega) \boldsymbol{\beta}^{H}(\psi, \omega) \boldsymbol{\Phi}\boldsymbol{N} }
 % where $\textbf{a} = vec({\rho} \mathbf{\alpha(\psi, \omega)} \mathbf{\beta}^{H}(\psi, \omega) \mathbf{\Phi} \mathbf{H} \mathbf{X} + \rho \boldsymbol{\alpha}(\psi, \omega) \boldsymbol{\beta}^{H}(\psi, \omega) \boldsymbol{\Phi}\boldsymbol{N})$ and
 % $\textbf{n}_{o} = \text{vec} (\textit{\textbf{N}}_{o}) \in \mathcal{CN}(0, \textbf{R}_{\textbf{n}_o})$ and $\textbf{R}_{\textbf{n}_o} = \sigma_{o}^{2} \textbf{I}_{N_{s}T_{o}}$.\\
 \small
\begin{subequations}
\begin{align}
&\textbf{b} = \text{vec}\left({\rho} \boldsymbol{\alpha}(\psi, \omega) \boldsymbol{\beta}^{H}(\psi, \omega) \boldsymbol{\Phi} \mathbf{H} \mathbf{X} + \rho \boldsymbol{\alpha}(\psi, \omega) \boldsymbol{\beta}^{H}(\psi, \omega) \boldsymbol{\Phi} \mathbf{N}_{a} \right), \\
&\textbf{n}_{o} = \text{vec}(\mathbf{N}_{o}) \in \mathcal{CN}(0, \mathbf{R}_{\mathbf{n}_o}), \quad \mathbf{R}_{\mathbf{n}_o} = \sigma_{o}^{2} \mathbf{I}_{N_{s}T}.
\end{align}
\end{subequations}
\normalsize
 Let the estimated parameters for target sensing $\tilde{\bm{\epsilon}} =$ [$\tilde{\bm{\xi}}^{T}$,$\textbf{$\tilde{\bm{\rho}}$}^{T}$]$^{T}$, where \textbf{$\tilde{\bm{\xi}}$} = [$\psi$,$\omega$]$^{T}$ and \textbf{$\tilde{\bm{\rho}}$} = [$\Re${$\{\rho\}$} + $\Im {\{\rho\}}]^{T}$.\\
 Hence, the fisher information matrix (FIM) for estimating parameter \textbf{$\tilde{\bm{\epsilon}}$} can be generally written as
 \begin{equation}
 \label{Jmatrix}
\mathbf{J} = \begin{bmatrix}
\mathbf{J_{\tilde{\bm{\xi}} \tilde{\bm{\xi}}}} & \mathbf{J_{\tilde{\bm{\xi}} \tilde{\bm{\rho}}}} \\
\mathbf{J_{\tilde{\bm{\xi}} \tilde{\bm{\rho}}}}^T & \mathbf{J_{\tilde{\bm{\rho}} \tilde{\bm{\rho}}}}
\end{bmatrix},
\end{equation}
where each element of \textbf{J} can be written as
\begin{align}
\label{Jij}
    \mathbf{J}_{i,j} & = \text{tr}\left(\mathbf{R}_{\mathbf{n}_o}^{-1} \frac{\partial \mathbf{R}_{\mathbf{n}_o}}{\partial \tilde{\mathbf{\tilde{\bm{\epsilon}}}}_i} \mathbf{R}_{\mathbf{n}_o}^{-1} \frac{\partial \mathbf{R}_{\mathbf{n}_o}}{\partial \tilde{\mathbf{\tilde{\bm{\epsilon}}}}_j}\right)\quad + 2 \Re\left\{\frac{\partial \textbf{b}^H}{\partial \tilde{\mathbf{\tilde{\bm{\epsilon}}}}_i} \mathbf{R}_{\mathbf{n}_o}^{-1} \frac{\partial \textbf{b}}{\partial \tilde{\mathbf{\tilde{\bm{\epsilon}}}}_j}\right\} \nonumber \\
    & \overset{\text{(a)}}{=} \frac{2}{\sigma_o^2} \Re\left\{\frac{\partial \textbf{b}^H}{\partial \tilde{\mathbf{\tilde{\bm{\epsilon}}}}_i} \frac{\partial \textbf{b}}{\partial \tilde{\mathbf{\tilde{\bm{\epsilon}}}}_j}\right\}.
\end{align}
This is because $\textbf{R}_{\textbf{n}_o}$ is independent of $\mathbf{\tilde{\bm{\epsilon}}}$ such 
that $\frac{\partial \textbf{R}_{\textbf{n}_o}}{\partial \tilde{\bm{\epsilon}_{i}}} = 0$ for any $i$. 
The derivation of the FIM elements is provided in Appendix A. Furthermore, the CRB for estimating $\tilde{\bm{\xi}}$ is given as
\begin{equation}
\text{CRB}(\tilde{\bm{\xi}}) = \left[ \textbf{J}_{\tilde{\bm{\xi}} \tilde{\bm{\xi}}} - \textbf{J}_{\tilde{\bm{\xi}} \tilde{\bm{\rho}}} \textbf{J}_{\tilde{\bm{\xi}} \tilde{\bm{\rho}}}^{-1} \textbf{J}_{\tilde{\bm{\xi}} \tilde{\bm{\rho}}}^T \right]^{-1}.
\end{equation}
Assuming that the target moves slowly, the target direction does not change remarkably over the adjacent coherent time slots. Accordingly, the predicted angles are enough for the waveform optimization that is necessary for minimizing the CRB. This is a typical scenario in radar tracking, where prior knowledge of the target direction is well-known for system design \cite{lyu2024crb}. Thus, the target angle $\mathbf{\tilde{\bm{\xi}}}$ is assumed to be fixed in this study.
\subsection{\textcolor{black}{Optimization Problem Formulation}}
The sum rate maximization problem can be formulated as
\begin{subequations}
\label{Optprob}
\begin{align}
\label{objective_function}
    \max_{\textbf{W}, \mathbf{\Phi}} & \quad \sum_{k=1}^K R_{k} \\
    \text{subject to:} \quad & \nonumber \\
    \label{constraint1}
    \gamma_{k} &\geq \gamma_{\text{th}}, \quad \forall k \in K, \\
    \label{constraint2}
    \text{tr}(\textbf{W} \textbf{W}^H) &\leq P_{\text{BS}}^{\text{max}}, \\
    \label{constraint3}
    E\left[\| \textbf{A} \mathbf{\Phi} (\textbf{H}\textbf{x} + \textbf{n}) \|^2\right] &\leq P_{\text{RIS}}^{\text{max}}, \\
    \label{constraint4}
    \| \textbf{g}_l^H \mathbf{A} \mathbf{\Phi} \|^2 \sigma_2^2 &\leq r_{\text{max}}, \\
    \label{constraint5}
    |\theta_i| &\leq 
    \begin{cases}
      1, & \forall i \notin \mathcal{A}, \\
      P_{A, \text{max}}, & \forall j \in \mathcal{A},
    \end{cases} \\
    \label{constraint6}
    \text{CRB}(\textbf{W}, \mathbf{\Phi}) &\leq \text{CRB}_{\text{max}}.
\end{align}
\end{subequations}
where (\ref{constraint1}) ensures the minimum SINR for communication users. (\ref{constraint2}) and (\ref{constraint3}) represent the total power budgets dedicated for the BS and the HRIS, respectively. (\ref{constraint4}) limits the thermal noise received at the the target within a certain range. (\ref{constraint5}) imposes an amplitude constraint on HRIS. (\ref{constraint6}) limits the CRB for the target estimation accuracy.
\section{DRL\textendash Based Joint Design of BS Beamforming and HRIS Precoding Matrix}
Since the objective function and
the constraints \textcolor{black}{in (\ref{Optprob})} are non\textendash convex leading to a non\textendash convex non\textendash trivial optimization
problem, obtaining the optimal solution by utilizing classical mathematical
tools would be impossible to achieve, specially for large scale network.
% Also, deploying alternating optimization
% techniques to maximize the objective function, where in each
% iteration, suboptimal $\mathbf{\Phi}$ is solved by first fixing $\textbf{W}$ while suboptimal $\mathbf{\Phi}$ is derived by fixing the $\textbf{W}$, until
% the algorithms converge. 
In our work, rather than directly
solving the challenging optimization problem mathematically,
we formulate the sum rate optimization problem in the context of DDPG\textendash based DRL method to obtain the feasible BS beamforming $\textbf{W}$ and the HRIS precoding matrix $\mathbf{\Phi}$.
\subsection{Proposed MDP}

The state and action spaces, and the reward that are used to represent our joint BS beamforming and HRIS precoding matrix optimization problem are designed as follows:
\begin{itemize}
\item{\textbf{State:}} The state vector $\mathbf{s}^{(t)}$ is composed of the current values of the BS beamforming matrix $\mathbf{W} \in \mathbb{C}^{M \times K+1}$, the current values of the elements in the main diagonal of the HRIS precoding matrix, i.e., in the vector $Diag(\bm{\Phi}) \in \mathbb{C}^{N \times 1}$, the elements in the matrix $\mathbf{H} \in \mathbb{C}^{N \times M}$ that stacks the channel gains from the BS to the HRIS, and the elements in the matrix $\textbf{G} \in \mathbb{C}^{K+1 \times N}$ that represent the channel gains from the HRIS to the $K$ users and the target such as $\textbf{G} = [\boldsymbol{g}_{1},...., \boldsymbol{g}_{K}, \boldsymbol{g}_{l}]$. Since the real and imaginary parts of complex-valued numbers can be treated as independent inputs, the actual dimension of the state space is $D_{\text{state}} = 2M (K+1) + 2N + 2NM + 2N(K+1)$. The state is constructed such that:
\begin{equation}
\begin{aligned}
\mathbf{s}^{(t)} = \{&\mathbf{H}^{(t)},\mathbf{G}^{(t)},\mathbf{W}^{(t)},diag(\bm{\Phi}^{(t)})\}.
\end{aligned}
\end{equation}
% \begin{equation}
% \begin{aligned}
% \mathbf{s}_t = [&\text{flatten}(\Re (\mathbf{H})), \text{flatten}(\Im (\mathbf{H})), \\
% &\text{flatten}(\Re (\mathbf{G})), \text{flatten}(\Im (\mathbf{G})), \\
% &\text{flatten}(\Re (\mathbf{W})), \text{flatten}(\Im (\mathbf{W})), \\
% &\Re (\mathbf{diag(\Phi)}), \Im (\mathbf{diag(\Phi)})],
% \end{aligned}
% \end{equation}
% where $\text{flatten}(\cdot)$ is the operator that transforms any matrix $\mathbf{V} \in \mathbb{C}^{A \times B}$ into a vector $\mathbf{v} \in \mathbb{C}^{1 \times AB}$, by first concatenating all the real parts and later the imaginary ones. Recall that the state space is continuous since the elements of $\bm{\theta}$ and $\mathbf{W}$ can take any real value as long as the optimization constraints in \ref{objfunc} are met.\\
\item \textbf{Action:} The action is simply constructed by the BS beamforming $\textbf{W}$ and the HRIS precoding matrix $\mathbf{\Phi}$. Likewise, to deal with the real input problem, $\textbf{W} = \Re{(\textbf{W})} + \Im{(\textbf{W})}$ and $\mathbf{\Phi} = \Re{(\mathbf{\Phi})} + \Im{(\mathbf{\Phi})}$ are separated as real and imaginary parts, both are entries of the action. Hence, the dimension of the action space is $D_{a} = 2M(K+1) + 2N$ and the action space is constructed such as:
\begin{equation}
\begin{aligned}
\mathbf{a}^{(t)} = \{&\mathbf{W}^{(t)},diag(\bm{\Phi}^{(t)})\}.
\end{aligned}
\end{equation}
% The action space is also continuous although its values are constrained to the real-valued interval $(-1, 1)$ in order to improve the system stability.
\item \textbf{Reward:} At the $t^{th}$ timestep of the DRL, the reward is determined as the sum rate $r(\textbf{H}^{(t)},\textbf{G}^{(t)},\textbf{W}^{(t)},diag(\bm{\Phi}^{(t)})$, given the instantaneous channels $\textbf{H}^{(t)}$ and $\textbf{G}^{(t)}$ and the action $\textbf{W}^{(t)}$ and $\mathbf{\Phi}^{(t)}$ obtained from the actor network.
\begin{algorithm}
\caption{DDPG for Joint BS Beamforming and HRIS Precoding Matrix Optimization}
\vspace{-0.5mm}
Initial state $s_0$, max episodes $E$, max steps per episode $T_{s}$, actor and critic networks, replay buffer $\mathcal{D}$, Network parameters (e.g. max amplitude $A_{max}$, max power $P_{max}$, BS Beamforming $\textbf{W}$, HRIS precoding matrix $\bm{\Phi}$, $\textbf{G}, \textbf{H}$,....etc);\\
% Initialize actor, critic, and target networks\;
Set initial state $s^{(t)} \leftarrow \text{normalize}(s_0^{(t)})$\;
\For{episode $i = 1$ to $E$}{
    Reset environment to initial state $s^{(t)} \leftarrow \text{normalize}(s_{0}^{(t)})$\;
    \For{timestep $t = 1$ to $T_{s}$}{
        Select action $a^{(t)}$ using actor network\;
        $a^{(t)} \leftarrow \text{scale\_actions}(a^{(t)}, A_{max}, P_{max})$\;
        Execute action $a_t$, observe reward $r_t$ and next state $s^{(t+1)}$\;
        $s^{(t+1)} \leftarrow \text{normalize}(s^{(t+1)})$\;
        Store transition $(s^{(t)}, a^{(t)}, r^{(t)}, s^{(t+1)})$ in replay buffer $\mathcal{D}$\;
        Sample mini-batch from replay buffer $\mathcal{D}$\;
        Update critic network by minimizing loss\;
        Update actor network using policy gradient\;
        Apply penalties if constraints in (\ref{Optprob}) are violated\;
        Set $s^{(t)} \leftarrow s^{(t+1)}$\;
    }
    \If{constraints in (\ref{Optprob}) are met or max timesteps reached}{
        End episode\;
    }
}
\Return Optimized BS beamforming matrix $\mathbf{W}_{opt}$ and HRIS precoding matrix $\bm{\Phi}_{opt}$\;
\end{algorithm}
 % \mathbf{a}_t$ is composed of variations in the elements of $\theta$ and $\mathbf{p}_k \, \forall k$ that compose the current state $\mathbf{s}_t$. Hence, the dimension of the action space is $D_{\text{action}} = 2K N_t + 2N$ and the action is constructed such that
    % \begin{equation}
    % \mathbf{a}_t = [\text{flatten}(\Delta \mathbf{p}_1), \ldots, \text{flatten}(\Delta \mathbf{p}_K), \text{flatten}(\Delta \theta)].
    % \end{equation}
\end{itemize}

\subsection{Algorithm}
In this sub\textendash section, the proposed DRL\textendash based algorithm for joint design of the BS beamforming and the HRIS precoding matrix is presented using the DDPG neural network. 
We assume there exists a central controller, or the agent, which is able to instantaneously collect the channel state information (CSI), $\mathbf{G}$ and $\mathbf{H}$. At time step $t$, given the CSI and the action $\mathbf{W}^{(t-1)}$ and $\mathbf{\Phi}^{(t-1)}$ in the previous state, the agent constructs the state $\mathbf{s}^{(t)}$ for time step $t$ following sub\textendash section III.B.

At the beginning of the algorithm, the experience replay buffer $\mathcal{D}$, the critic network and the actor network parameters, the action $\mathbf{W}$ and $\mathbf{\Phi}$ need to be initialized. In this paper, we simply adopt the identity matrix to initialize $\mathbf{W}$ and $\mathbf{\Phi}$.

Our algorithm is run over $E$ episodes, where every episode iterates $T_{s}$ steps. For each episode, the algorithm terminates whenever constraints (\ref{constraint1}), (\ref{constraint2}), (\ref{constraint3}), (\ref{constraint4}) and (\ref{constraint5}), (\ref{constraint6}) are met or the algorithm reaches the maximum number of allowable time steps per episode. The optimal BS beamforming $\mathbf{W}_{opt}$ and HRIS precoding $\mathbf{\Phi}_{opt}$ are obtained as the action with the best instant reward. The details of the proposed method are shown in \textbf{Algorithm 1}.
% \vspace{-5mm}
\SetNlSty{}{}{}  % Remove style for line numbers
\SetAlgoNlRelativeSize{-1}  % Reduce the size of line numbers
\SetInd{0.3em}{0.3em}  % Reduce indentation
\SetEndCharOfAlgoLine{}  % Remove end of line markers
\setlength{\algomargin}{1.5em}  % Reduce margin
\SetAlgoNlRelativeSize{-1}  % Further reduce line number size
% \vspace{-10mm}
\begin{table}
\vspace{-15mm}
\caption{Simulation Parameters.}
        \begin{center}
      \renewcommand{\arraystretch}{1}
 \begin{tabular}{lll}
            \toprule
            % \addlinespace
            Parameter & Description & Value \\
            \hline
            \hline
            % \addlinespace
         $E$ & Number of Episodes & $10$ \\

        $n_{s}$ & Number of time steps per episode & $100$ \\
                $B$ & Min\textendash batch size & $100$ \\
         $\gamma_{d}$ & Discount factor & $0.99$ \\
                  $\gamma_{d}$ & Soft update rate for target networks & $0.005$ \\
        $lr$ & Learning rate &  $10^{-4}$\\
  $n_{1}$, $n_{2}$ &Neural Network Dimension &\{64,64\}\\      
          $f$ & Operating frequency &  $0.2$ THz\\
            $k(f)$ & Absorption coefficient & $0.01$\\
              $M$ & Number of BS antennas & 64\\
               $N$ &  Number of RIS reflecting elements & 80\\
                              $q$ &  Number of RIS active elements & 30\\
               $N_{s}$ &  Number of RIS sensing elements & 20\\
                $K$ & Number of users & $3$ \\
                $l$ & Number of Targets & $1$ \\
                $A_{max}$& Maximum amplitude of active RIS elements & 5\\
                 $CRB_{max}$ & Maximum CRB allowed & $10^{-3}$\\
             $\sigma_{o},\sigma_{a}^{2}$ & Noise variance & $-90$ dBm\\
      $P_{\text{BS}}^{\text{max}}$ & BS Power Budget & $30$ dBm\\
       $P_{\text{RIS}}^{\text{max}}$ & RIS Power Budget & $10$ dBm\\
             \bottomrule
         \end{tabular}
 \renewcommand{\arraystretch}{1}
         \end{center}
\label{tab:capacity}
\end{table}
\section{Simulations Results}
In this section, we aim to demonstrate the performance of the proposed DDPG approach for jointly optimizing the BS beamforming and the HRIS precoding matrix for the sake of maximizing the sum rate. The environment settings and model parameters used in these simulations are detailed in Table 1.\\
% In our system, a 64\textendash antenna BS transmit signals to three single\textendash antenna users, randomly located at a fixed distance of 10 m from the RIS. The BS and HRIS power budgets are limited to $30$ dBm and $10$ dBm, respectively. The RIS has a total of 100 elements divided between 81 reflecting elements and 19 sensing elements, where the HRIS sensing elements are capable of sensing one target.\\
To verify the effectiveness of our proposed scheme, we select four benchmark algorithms: random BS beamforming and HRIS phase  shifts, greedy algorithm, PPO and SAC. These benchmarks provide a comprehensive comparison across different approaches to the joint BS beamforming and HRIS precoding matrix optimization. The random BS beamforming and HRIS precoding matrix scheme serves as a non\textendash optimized baseline, while greedy algorithm represents an efficient heuristic based algorithm that may not reach globally optimal results. The PPO and SAC algorithms are included for direct comparisons as other ML baselines.
\begin{figure}[ht]
\centering
\vspace{-0.2 in}
\includegraphics[height=5cm,width=6cm]{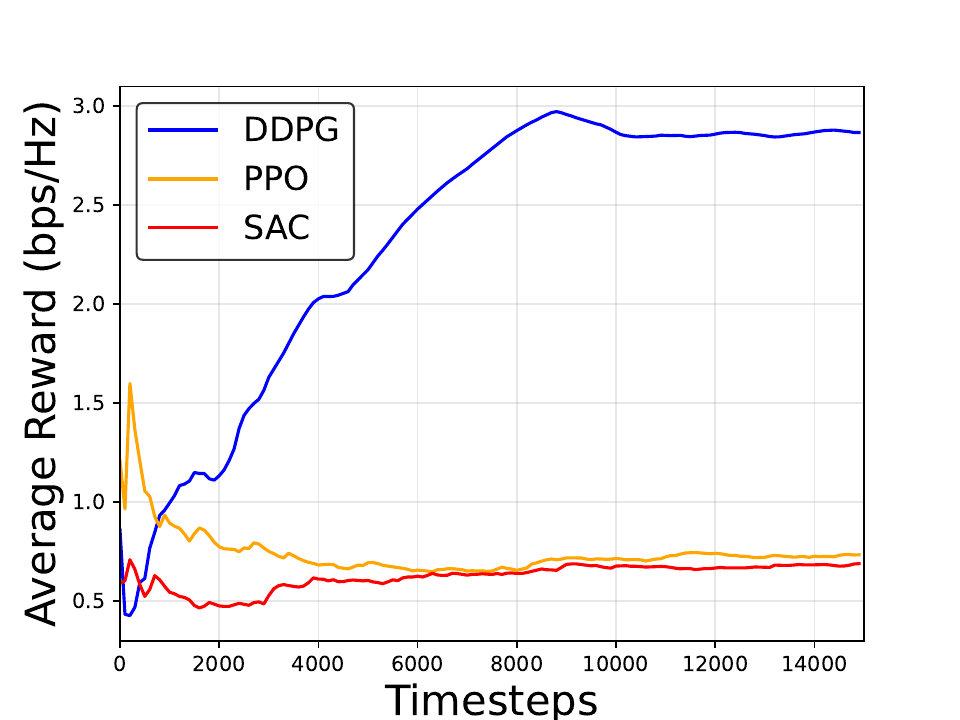}
\caption{Average Reward vs. Timesteps for Different Algorithms.}\label{convergence}
 \vspace{-0.01 in}
\end{figure}

\begin{figure}[ht]
\centering
\vspace{-0.2 in}
\includegraphics[height=5cm,width=6cm]{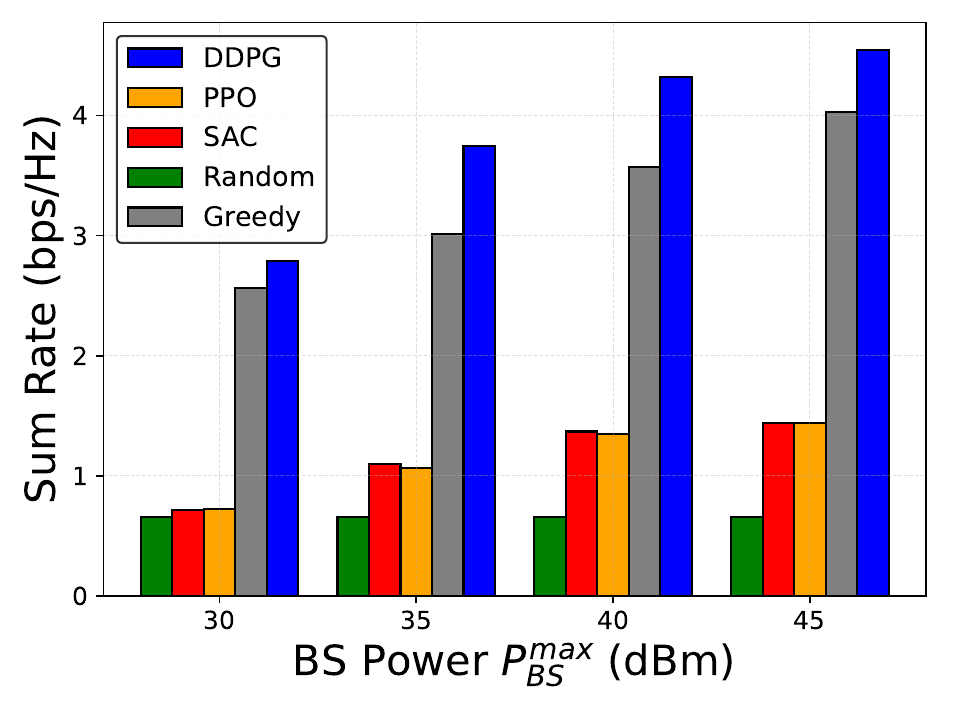}
\caption{Average Sum Rate vs. BS Power (dBm).}
% \notered{Increase the size of the x-axis and y-axis fonts, legends. Also, why is PPO and SAC are so bad? Was that their best performance even if you selected the best hyper parameters? }\textcolor{orange}{yes, this was the case}}
\label{power}
 \vspace{-0.01 in}
% \vspace{-2 mm}
\end{figure}

\begin{figure}[ht]
\centering
\vspace{-2 mm}
\includegraphics[height=5cm,width=6cm]{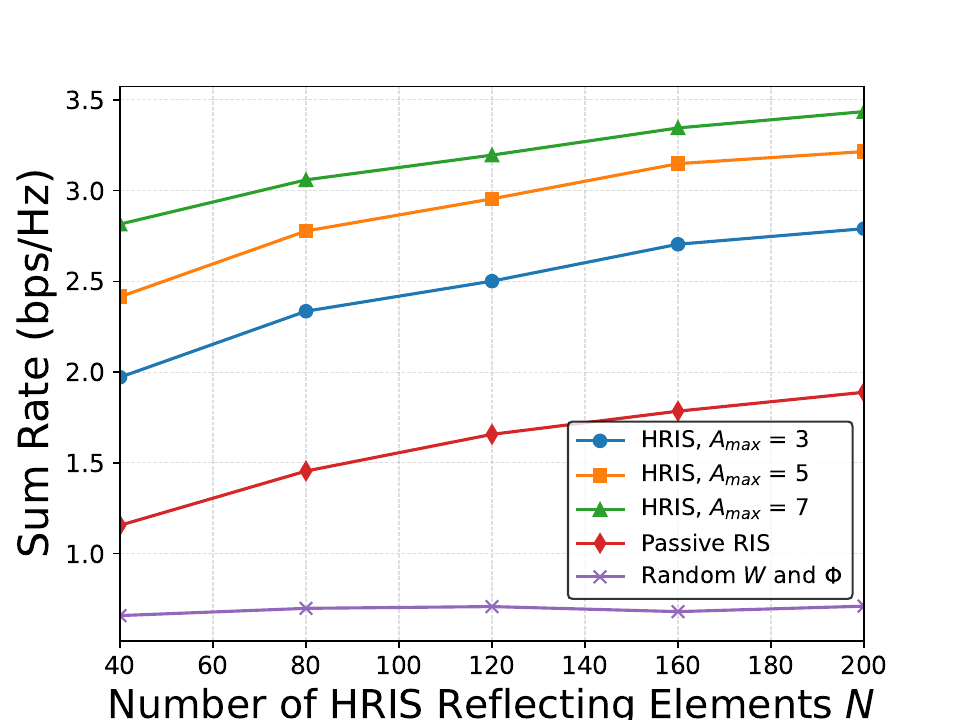}
\vspace{-2 mm}
\caption{Average Sum Rate vs. the Number of HRIS reflecting elements for Passive RIS, Random BS Beamforming $\textbf{W}$ and HRIS Precoding Matrix $\bm{\Phi}$ schemes and the proposed HRIS scheme at different values of maximum amplitude of active HRIS elements $A_{max}$.}\label{RISelements}
% \vspace{-0.1 in}
\end{figure}

Fig.\ref{convergence} shows the convergence of the average reward over iterations for DDPG, PPO and SAC. The three algorithms successfully converge, but DDPG consistently achieves the highest rewards outperforming other ML algorithms. This proves the effectiveness and superiority of optimizing the BS bemforming and HRIS precoding matrix using the DDPG in such \textcolor{black}{SS\textendash} HRIS-assisted ISAC scenario.

Fig.\ref{power} shows the sum rate as a function of the BS power budget for various algorithms. As the BS power budget increases, DDPG consistently achieves the highest sum rate proving its robustness in improving the sum rate under varying power levels, followed by the greedy algorithm which still performs well but lags DDPG. At low BS power values, such as 30 dBm,  PPO, SAC and random approach perform relatively similarly. As the BS power increases further, both SAC and PPO outperforms the random scheme indicating the importance of structured decision-making for better performance.

Fig.\ref{RISelements} illustrates the effect of the number of reflecting elements $N$ on the sum rates of different schemes with BS power $P_{max} = 30$ dBm and RIS sensing elements $N_{s} = 20$ using DDPG, where the number of active RIS elements is set as $\left\lceil \frac{N}{4} \right\rceil$. It is readily observed that for both the HRIS and the passive RIS, the sum rates increase with $N$ thanks to the enhanced spatial degrees of freedom (DoFs) of RIS. Additionally, the HRIS schemes yields a significantly higher achievable sum rate compared to the passive RIS scheme, especially at higher values of maximum amplitude of active HRIS elements \textcolor{black}{$A_{max}$}. This is owing to the HRIS capabilities of providing power amplification gain and passive beamforming gain simultaneously. However, due to the limited amplification power at the HRIS, it is noticed that the sum rate achieved by the HRIS scheme increases slowly when $N$ becomes relatively large.

\section{Conclusions}
In this paper, we adapted an \textcolor{black}{SS\textendash} HRIS downlink scenario, where the HRIS is capable of both reflecting incident signals as well as sensing the received radar echo signal from a target. The joint BS beamforming and HRIS precoding matrix optimization problem is proposed for the sake of maximizing 
the sum rate of communication users while guaranteeing the target accuracy estimation measured by the CRB and thermal noise. Our optimization problem is solved using the policy\textendash based DDPG derived from Marcov decision process to optimize continuous BS beamforming matrix and HRIS precoding matrix. Through comprehensive simulations, we have demonstrated that DDPG significantly outperforms other benchmark algorithms, such as PPO, SAC, Greedy and random algorithms. Additionally, our analysis confirmed that HRIS when combined with an increased number of RIS elements, provides substantial gains compared to passive RIS and random BS beamforming and HRIS precoding matrix schemes.
\section*{Acknowledgment}
\vspace{-2mm}
This work has been supported by NSERC Canada Research Chairs program.
\begin{figure*}[t]
\small
\begin{align}
 \label{Jww}
    &\mathbf{J}_{\tilde{\bm{\xi}}\tilde{\bm{\xi}}} = \frac{2 | \rho |^2}{\sigma_o^2} \Re \left\{
    \begin{bmatrix}
    T \cdot \text{Tr}(\hat{\bm{\Omega}}_{\psi} \bm{\Phi} \bm{H} \bm{R_x} \bm{H}^H \bm{\Phi}^H \hat{\bm{\Omega}}_{\psi}^H) + \sigma_{a}^{2} \cdot \text{Tr}(\hat{\bm{\Omega}}_{\psi} \bm{\Phi} \bm{\Phi}^H \hat{\bm{\Omega}}_{\psi}^H) &
    T \cdot \text{Tr}(\hat{\bm{\Omega}}_{\psi} \bm{\Phi H} \bm{R_x} \bm{H}^H \bm{\Phi}^H \hat{\bm{\Omega}}_{\omega}^H) +  \sigma_{a}^{2}  \cdot \text{Tr}(\hat{\bm{\Omega}}_{\psi} \bm{\Phi} \bm{\Phi}^H \hat{\bm{\Omega}}_{\omega}^H) \\
    T \cdot \text{Tr}(\hat{\bm{\Omega}}_{\omega} \bm{\Phi} \bm{H} \bm{R_x} \bm{H}^H \bm{\Phi}^H \hat{\bm{\Omega}}_{\psi}^H) +  \sigma_{a}^{2}  \cdot \text{Tr}(\hat{\bm{\Omega}}_{\omega} \bm{\Phi} \bm{\Phi}^H \hat{\bm{\Omega}}_{\psi}^H) &
    T \cdot \text{Tr}(\hat{\bm{\Omega}}_{\omega} \bm{\Phi} \bm{H} \bm{R_x} \bm{H}^H \bm{\Phi}^H \hat{\bm{\Omega}}_{\omega}^H) +  \sigma_{a}^{2}  \cdot \text{Tr}(\hat{\bm{\Omega}}_{\omega} \bm{\Phi} \bm{\Phi}^H \hat{\bm{\Omega}}_{\omega}^H)
    \end{bmatrix}
    \right\}.
    \end{align}
    \vspace{-0.5in}
 %   \hrulefill
\end{figure*}
% \begin{equation}
% \begin{aligned}
% \label{Jpsirho}
% \mathbf{J}_{\tilde{\xi}\tilde{\rho}} &= \frac{2}{\sigma_o^2} \Re \left\{  
%      \left[\begin{matrix}
% \rho^* \text{vec} {(\hat{\mathbf{\Omega}}_\psi \Phi \mathbf{H} \mathbf{X}+ \hat{\mathbf{\Omega}}_\psi \Phi \mathbf{N})}^{H} \\
% \rho^* \text{vec} {(\hat{\mathbf{\Omega}}_\omega \Phi \mathbf{H} \mathbf{X}+ \hat{\mathbf{\Omega}}_\omega \Phi \mathbf{N})}^{H}
% \end{matrix} \right] \left\{ \left[ 1, j \right] \otimes vec{(\mathbf{\Omega} \Phi \mathbf{H} \mathbf{X}+ \mathbf{\Omega} \Phi \mathbf{N})} \right\}  \right\}\\
%     &= \frac{2}{\sigma_o^2} \Re \left\{ 
% M_{ij} \otimes \left( 
% \rho^* \text{Tr}\left(\hat{\Omega}_{\psi} \Phi H R_x H^H \Phi^H \Omega^H \right) + \rho \text{Tr}\left( \hat{\Omega}_{\omega} \Phi H R_x H^H \Phi^H \Omega^H \right)
% + \rho^* \text{Tr}\left( \hat{\Omega}_{\psi} \Phi n_x \Phi^H \Omega^H \right) + \rho \text{Tr}\left( \hat{\Omega}_{\omega} \Phi n_x \Phi^H \Omega^H \right)
% \right) 
% \right\}
% \end{aligned}
% \end{equation}
\begin{figure*}
\begin{align}
 \label{Jpsirho}
\mathbf{J}_{\tilde{\bm{\xi}}\tilde{\bm{\rho}}} 
&= \frac{2}{\sigma_o^2} \Re \left\{ 
 \left( [1,j]^{H} [1,j] \right)  \otimes \left( 
 T \rho^* \text{Tr}\left(\hat{\bm{\Omega}}_{\psi} \bm{\Phi} \bm{H} \bm{R_x} \bm{H}^H \bm{\Phi}^H \bm{\Omega}^H \right) 
+  T \rho \text{Tr}\left( \hat{\bm{\Omega}}_{\omega} \bm{\Phi} \bm{H R_x} \bm{H}^H \bm{\Phi}^H \bm{\Omega}^H \right) \right. \right. \nonumber\\
& \quad \left. \left. 
+\sigma_{a}^{2} \rho^* \text{Tr}\left( \hat{\bm{\Omega}}_{\psi} \bm{\Phi} \bm{\Phi}^H \bm{\Omega}^H \right) 
+\sigma_{a}^{2} \rho \text{Tr}\left( \hat{\bm{\Omega}}_{\omega} \bm{\Phi} \bm{\Phi}^H \bm{\Omega}^H \right)
\right) 
\right\}.
\end{align}
\vspace{-0.3in}
\end{figure*}
\begin{figure*}
\begin{align}
 \label{Jrhorho}
    \mathbf{J}_{\tilde{\rho} \tilde{\rho}} = \frac{2}{\sigma_o^2} \Re \left\{ \left( [1,j]^{H} [1,j] \right) \otimes \left( T \cdot \text{Tr}(\bm{\Omega} \bm{\Phi} \bm{H R_x H}^H \bm{\Phi}^H \bm{\Omega}^H) + \sigma_{a}^{2} \cdot\text{Tr}(\bm{\Omega \Phi \Phi}^H \bm{\Omega}^H) \right) \right\}.
    \end{align}
    \hrulefill
\end{figure*}
\appendix
\section*{Appendix A: Derivation of Fisher Information Matrix for the Point Target}
From (\ref{Jij}), the partial derivatives of $\mathbf{b}$ with respect to $\tilde{\bm{\xi}}$ and $\tilde{\bm{\rho}}$ are expressed as follows:
\small
\begin{equation}
    \frac{\partial \textbf{b}}{\partial \tilde{\mathbf{\tilde{\bm{\xi}}}}} = \left[ \rho vec{(\hat{\mathbf{\Omega}}_\psi \Phi \mathbf{H} \mathbf{X} + \hat{\mathbf{\Omega}}_\psi \Phi \mathbf{N}_{a})}, \rho vec{(\hat{\mathbf{\Omega}}_\omega \Phi \mathbf{H} \mathbf{X}+ \hat{\mathbf{\Omega}}_\omega \Phi \mathbf{N}_{a})} \right]
\end{equation}
\normalsize
\begin{equation}
    \frac{\partial \textbf{b}}{\partial \tilde{\mathbf{\bm{\rho}}}} = \left[ 1, j \right] \otimes vec{(\mathbf{\Omega} \Phi \mathbf{H} \mathbf{X}+ \mathbf{\Omega} \Phi \mathbf{N}_{a})} 
\end{equation}
where $\mathbf{\Omega} = \mathbf{\bm{\alpha}}(\psi, \omega) \mathbf{\bm{\beta}}^T(\psi, \omega)$. In order to derive the partial derivatives of $\mathbf{\Omega}$ with respect to $\psi$ and $\omega$ (denoted as $\hat{\mathbf{\Omega}}_\psi$ and $\hat{\mathbf{\Omega}}_\omega$), the steering vectors can be rewritten as:
\begin{equation}
\mathbf{\bm{\alpha}}(\psi, \omega) = \frac{1}{\sqrt{N_s}} e^{-j\bm{\delta}_s}, \quad \mathbf{\bm{\beta}}(\psi, \omega) = \frac{1}{\sqrt{N}} e^{-j\bm{\delta}}
\end{equation}
where
\begin{align}
\bm{\delta}_s &= \left[ \frac{2\pi}{\lambda} \right] (\kappa_{sY} \sin \psi \sin \omega d_y + \kappa_{sZ} \cos \omega d_z), \\
\bm{\delta} &= \left[ \frac{2\pi}{\lambda} \right] (\kappa_Y \sin \psi \sin \omega d_y + \kappa_Z \cos \omega d_z)
\end{align}
where $\kappa_{sY}$ and $\kappa_{sZ}$ denote the
element indices of sensing elements at Y and Z axes, respectively,
and $\kappa_{Y}$ and $\kappa_{Z}$ denote those of reflecting elements.
Hence, the partial derivatives of $\mathbf{\Omega}$ can be written as:
\small
\begin{align*}
\hat{\mathbf{\Omega}}_\psi= \frac{-2j\pi}{\lambda \sqrt{N_s N}} \cos \psi \sin \omega d_y\left( \text{diag}\{\kappa_{sY}\} \mathbf{\bm{\alpha}} \mathbf{\bm{\beta}}^T + \mathbf{\bm{\alpha}} \mathbf{\bm{\beta}}^T \text{diag}\{\kappa_Y\} \right)
\end{align*}
\begin{align*}
\hat{\mathbf{\Omega}}_\omega =& \frac{-2j\pi}{\lambda \sqrt{N_s N}} \sin \psi \cos \omega d_y\left( \text{diag}\{\kappa_{sY}\} \mathbf{\bm{\alpha}} \mathbf{\bm{\beta}}^T + \mathbf{\bm{\alpha}} \mathbf{\bm{\beta}}^T \text{diag}\{\kappa_Y\} \right) \\
\quad + & j \frac{2\pi}{\lambda \sqrt{N_s N}} \sin \psi d_z \left( \text{diag}\{\kappa_{sZ}\} \mathbf{\bm{\alpha}} \mathbf{\bm{\beta}}^T + \mathbf{\bm{\alpha}} \mathbf{\bm{\beta}}^T \text{diag}\{\kappa_Z\} \right)
\end{align*}
\normalsize
% \begin{align*}
% &\hat{\mathbf{\Omega}}_\psi = \frac{\partial \mathbf{\Omega}}{\partial \psi} = \frac{\partial \mathbf{\bm{\alpha}}(\psi, \omega)}{\partial \psi} \mathbf{\bm{\beta}}^T(\psi, \omega) + \mathbf{\bm{\alpha}}(\psi, \omega) \frac{\partial \mathbf{\bm{\beta}}^T(\psi, \omega)}{\partial \psi} \\
% &= \frac{-2j\pi}{\lambda \sqrt{N_s N}} \cos \psi \sin \omega d_y\quad \times \left( \text{diag}\{\kappa_{sY}\} \mathbf{\bm{\alpha}} \mathbf{\bm{\beta}}^T + \mathbf{\bm{\alpha}} \mathbf{\bm{\beta}}^T \text{diag}\{\kappa_Y\} \right)
% \end{align*}
% \begin{align*}
% &\hat{\mathbf{\Omega}}_\omega = \frac{\partial \mathbf{\Omega}}{\partial \omega} = \frac{\partial \mathbf{\alpha}(\psi, \omega)}{\partial \omega} \mathbf{\beta}^T(\psi, \omega) + \mathbf{\bm{\alpha}}(\psi, \omega) \frac{\partial \mathbf{\bm{\beta}}^T(\psi, \omega)}{\partial \phi} \\
% &= \frac{-2j\pi}{\lambda \sqrt{N_s N}} \sin \psi \cos \omega d_y\quad \times \left( \text{diag}\{\kappa_{sY}\} \mathbf{\bm{\alpha}} \mathbf{\bm{\beta}}^T + \mathbf{\bm{\alpha}} \mathbf{\bm{\beta}}^T \text{diag}\{\kappa_Y\} \right) \\
% &\quad + j \frac{2\pi}{\lambda \sqrt{N_s N}} \sin \psi d_z \quad \times \left( \text{diag}\{\kappa_{sZ}\} \mathbf{\bm{\alpha}} \mathbf{\bm{\beta}}^T + \mathbf{\bm{\alpha}} \mathbf{\bm{\beta}}^T \text{diag}\{\kappa_Z\} \right)
% \end{align*} 
With $\hat{\mathbf{\Omega}}_\psi$ and $\hat{\mathbf{\Omega}}_\omega$, the FIM elements $\mathbf{J}_{\tilde{\bm{\xi}}\tilde{\bm{\xi}}}$, $\mathbf{J}_{\tilde{\bm{\xi}}\tilde{\bm{\rho}}}$ and $\mathbf{J}_{\tilde{\bm{\rho}}\tilde{\bm{\rho}}}$ are expressed in \textcolor{black}{(\ref{Jww}), (\ref{Jpsirho}) and (\ref{Jrhorho}), respectively, at the top of the page.}
%%%I need to modify the numbering and make sure the equations are really at thr top of the page (final step).
% Accordingly, each entry of eqn.(\ref{Jmatrix}) is computed by eqns. (\ref{Jww}), (\ref{Jpsirho}) and (\ref{Jrhorho}).\\
\bibliographystyle{IEEEtran}
\bibliography{bibliography}
\end{document}